\documentclass[iop,apj]{emulateapj}
\usepackage{txfonts}

\usepackage{graphicx}

\usepackage[protrusion=true]{microtype}

\usepackage{color}
\definecolor{cite}{rgb}{0.0,0.0,0.8}
\usepackage[colorlinks=true,citecolor=cite,urlcolor=cite]{hyperref}

\usepackage{booktabs}

\def\ls{LS\,5039}
\def\lsi{LS\,I\,+61\,303}
\def\psr{PSR\,B1259$-$63}
\newcommand{\ergcms}{\ensuremath{\mathrm{erg\,cm^{-2}\,s^{-1}}}}
\newcommand{\ergs}{\ensuremath{\mathrm{erg\,s^{-1}}}}
\newcommand{\msolyr}{\ensuremath{\mathrm{M_\odot\,yr^{-1}}}}
\newcommand{\msol}{\ensuremath{\mathrm{M_\odot}}}
\newcommand{\lsd}{\ensuremath{L_\mathrm{sd}}}
\renewcommand{\email}[1]{\href{mailto:#1}{#1}}

\newlength\figwidth
\shorttitle{THERMAL X-RAY EMISSION FROM THE SHOCKED STELLAR WIND OF
PULSAR GAMMA-RAY BINARIES}
\shortauthors{Zabalza, Bosch-Ramon, \& Paredes}
\submitted{Accepted 2011 August 18}

\begin{document}

\title{Thermal X-ray emission from the shocked stellar wind 
        of pulsar gamma-ray binaries}

\author{V. Zabalza$^1$, V. Bosch-Ramon$^2$, and J.M. Paredes$^1$}

\affil{$^1$ Departament d'Astronomia i Meteorologia, Institut de Ci\`encies
del Cosmos (ICC), Universitat de Barcelona (IEEC-UB),\\
Mart\'i i Franqu\`es, 1, E08028, Barcelona, Spain;
\email{vzabalza@am.ub.es}} 
\affil{$^2$ Dublin Institute for Advanced Studies,
31 Fitzwilliam Place, Dublin 2, Ireland.}

\begin{abstract}
Gamma-ray loud X-ray binaries are binary systems that show non-thermal broadband
emission from radio to gamma rays. If the system comprises a massive star and a
young non-accreting pulsar, their winds will collide producing broadband
non-thermal emission, most likely originated in the shocked pulsar wind.
Thermal X-ray emission is expected from the shocked stellar wind, but until now
it has neither been detected nor studied in the context of gamma-ray binaries.
We present a semi-analytic model of the thermal X-ray emission from the shocked
stellar wind in pulsar gamma-ray binaries, and find that the thermal X-ray
emission increases monotonically with the pulsar spin-down luminosity, reaching
luminosities of the order of $10^{33}\,\ergs$.  The lack of thermal features in
the X-ray spectrum of gamma-ray binaries can then be used to constrain the
properties of the pulsar and stellar winds. By fitting the observed X-ray
spectra of gamma-ray binaries with a source model composed of an absorbed
non-thermal power law and the computed thermal X-ray emission, we are able to
derive upper limits on the spin-down luminosity of the putative pulsar. We
applied this method to \ls, the only gamma-ray binary with a radial, powerful
wind, and obtain an upper limit on the pulsar spin-down luminosity of $\sim
6\times10^{36}\,\ergs$.  Given the energetic constraints from its high-energy
gamma-ray emission, a non-thermal to spin-down luminosity ratio very close to
unity may be required.
\end{abstract}


\section{Introduction}\label{sec:intro}


In the past few years, gamma-ray binaries have emerged as a new category
of very high-energy (VHE, $E>100$~GeV) gamma-ray sources. In most of the
cases, they comprise a stellar mass compact object orbiting a young,
massive star. They generally differ from typical X-ray binaries in that
they exhibit luminosities at gamma-ray energies at levels similar or
above their X-ray luminosity. 
Five such sources have been already detected:
\ls\ \citep{2005Sci...309..746A}, \psr\ \citep{2005A&A...442....1A}, \lsi\
\citep{2006Sci...312.1771A}, HESS\,J0632$+$057
\citep{2009MNRAS.399..317S,
2011ApJ...737L..11B,2011ATel.3180....1M}, and
1FGL\,J1018.6$-$5856 
\citep{2011ATel.3221....1C,2011ATel3228....1P}.  Evidence of VHE flaring
emission has been found in Cygnus~X-1 \citep{2007ApJ...665L..51A}, but
at present lacks confirmation.

Whereas the pulsar nature of the compact object in \psr\ is determined thanks to
the detection of radio pulses, this is not possible in the case of \ls\ and \lsi, 
in which the radio pulses would be likely free--free absorbed in the stellar wind. 
Broadband radiation modeling in the microquasar scenario has been able to
reproduce the spectral energy distribution of the source
\citep[e.g.,][]{2006A&A...451..259P,2006A&A...459L..25B}, but the lack of
accretion signatures, and the similarities with \psr, prompted other authors to
consider the pulsar wind shock (PWS) scenario for these sources
\citep[e.g.,][]{2005A&A...430..245M,2006A&A...456..801D,2006MNRAS.372.1585C}.
%

The emission from the non-thermal particle population accelerated in the PWS
scenario has been widely studied in the radio, X-ray, and high energy
(HE) and VHE gamma-ray
bands 
\citep{1981MNRAS.194P...1M, 1997ApJ...477..439T, 1999APh....10...31K,
2006A&A...456..801D,2007MNRAS.380..320K,2007Ap&SS.309..253N,2011arXiv1103.2996B}.
Furthermore, the possibility of detecting the free pulsar wind from its spectral
signature has also been discussed \citep[e.g.,][]{1999APh....10...31K,
2007MNRAS.380..320K, 2009A&A...507.1217C}. However, one of the sources of
emission that could provide valuable information on the characteristics of
gamma-ray binaries has been hitherto overlooked: the thermal X-ray emission
from the shocked stellar wind.  Massive stellar binaries (of types O+O or O+WR)
are known to produce powerful thermal X-ray emission in the wind collision
region (WCR), where the supersonic winds of the two young stars interact
\cite[see, e.g.,][for a review]{2005xrrc.procE2.01P}. Even though it has been
previously considered that the detection of X-ray emission lines would be an
indication of accretion in gamma-ray binaries,
similar lines can arise from the thermal
X-ray emission of the shocked stellar wind in the WCR. The complexity of the
stellar winds from the Be companion stars in \lsi\ and \psr\ makes the analysis
of the thermal emission from the shocked stellar wind difficult. The presence of
both a polar wind and a decretion disk adds a number of unknown free
parameters. On the other hand, the radial wind from the O star in \ls, also more
powerful than the one from Be stars, turns this source into an ideal candidate
to study the observational impact of the thermal wind emission.


In this paper we present a model for the thermal X-rays from pulsar
gamma-ray binaries inspired on semi-analytical models of the X-ray
emission from massive binary systems
\citep[e.g.,][]{2004ApJ...611..434A,2008MNRAS.388.1047P}.  This model,
which allows the study of the dependence of the thermal X-ray emission
for different stellar and pulsar wind properties, will be applied to the
binary \ls\ in order to constrain the spin-down luminosity of its
putative pulsar. In Section~{\ref{sec:dynamic}}, we characterize the
shocked stellar wind region, approximated here as the contact
discontinuity between the pulsar and the stellar winds, and present a
model of its thermal X-ray emission in Section~\ref{sec:thermal}. In
Section~\ref{sec:appls} we apply the model to \ls\ and present 
spin-down luminosity upper limits in Section~\ref{sec:results}.
Finally, we discuss the implications of the results in
Section~\ref{sec:discussion}.


\section{The dynamical model}\label{sec:dynamic}


In the PWS scenario, the pulsar wind and the stellar wind collide forming an
interaction region bounded on either side by the reverse shocks of the winds.
The shocked winds are separated by a surface called contact discontinuity (CD),
located where the ram pressure components perpendicular to this surface are in
equilibrium: $p_{\star\perp}=p_{p\perp}$. Both winds are assumed here to be
radial.

The ram pressure of the relativistic pulsar wind can be represented at any point
at a distance $r_p$ from the pulsar as $p_p(r_p)=\lsd/(4\pi c r^2_p)$, where
$\lsd$ is the spin-down luminosity of the pulsar. In the case of the stellar
wind, we will assume a $\beta$-velocity law in the radial direction,
$v(r_\star)=v_\infty(1-{R_\star}/{r_\star})\,^\beta$, where $r_\star$ is the
distance to the center of the star, $R_\star$ is the radius of the star and
$v_\infty$ is the terminal velocity of the stellar wind. The azimuthal component
of the wind will be neglected.
We have considered $\beta=0.8$, which has been shown to be a good
approximation of the outer structure of the winds of hot massive stars
\citep{1986A&A...164...86P}. Assuming a constant mass-loss rate from the star
$\dot{M}$, the density of the stellar wind is $\rho=\dot{M}/(4\pi r^2_\star
v(r_\star))$.  The ram pressure of the stellar wind is given by
$p_\star(r_\star)=\dot{M}v(r_\star)/(4\pi r^2_\star)$. 

\begin{figure}
    \includegraphics[width=\figwidth]{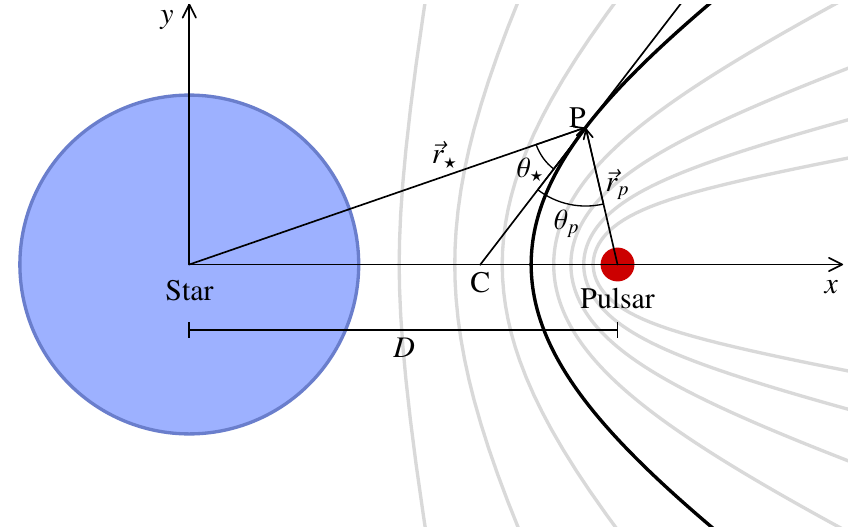}
    \caption{Illustration of the dynamical model for a close pulsar gamma-ray
    binary. The $\sim$10\,R$_\odot$ star (but not the pulsar) is at
    scale. The thick black line indicates the shape of the CD for
    $\eta_\infty=0.04$. The line \emph{CP} is tangent to the CD at the point
    \emph{P}, for which the angles and radii used in the derivation of the CD
    are shown.  The gray lines show the shape of the CD for values of
    $\eta_\infty$ spaced logarithmically in the range $0.003\mbox{--}0.3$.}
    \label{fig:shape}
\end{figure}

The balance of perpendicular components of the ram pressure that defines the
position of the CD
\begin{equation}
    \frac{\dot{M}v(r_\star)}{4\pi r^2_\star}\sin^2\theta_\star=
    \frac{\lsd/c}{4\pi r^2_p}\sin^2\theta_p,
    \label{eq:balance}
\end{equation} 
can be rearranged to obtain the dimensionless parameter $\eta$
\begin{equation}
    \frac{r^2_p\sin^2\theta_\star}{r^2_\star\sin^2\theta_p} = 
    \frac{\lsd/c}{\dot{M}v(r_\star)} \equiv
    \eta(x,y),
    \label{eq:eta}
\end{equation}
where $\theta_\star$ and $\theta_p$ are the angles between the CD surface and
the directions toward the center of the star and the pulsar, respectively. The
coordinates $x$ and $y$ correspond to the position along the line of centers and
the distance from it, respectively (see Figure~\ref{fig:shape}).  Following
\cite{2004ApJ...611..434A}, we can obtain the differential equation that
describes the shape of the CD as a function of $\eta$ from
Equation~(\ref{eq:eta}),
\begin{equation} 
    \frac{\mbox{d}x}{\mbox{d}y}=
    \frac{1}{y}\left[x-\frac{Dr_\star^2(x,y)\sqrt{\eta(x,y)}}{r^2_\star(x,y)\sqrt{\eta(x,y)}+r^2_p(x,y)}\right],
    \label{eq:difeq}
\end{equation}
where $x(y)$ is the function describing the shape of the CD and $D$ is the
binary separation distance. The boundary condition
($\mathrm{d}x/\mathrm{d}y\,\vline_{\,y\rightarrow0}\rightarrow0$) is
$x_0=D/(1+\sqrt{\eta})$, the location of pressure equilibrium along the symmetry
axis, also known as stagnation radius. For the case of constant wind velocities,
$\eta$ will be constant, and $x_0$ and $x(y)$ can be computed analytically.
However, given the $\beta$-velocity law we adopted for the stellar wind, a
numerical integration is required to obtain the shape of the CD since the
stellar wind may not yet have reached its terminal velocity at the contact
surface. Figure~\ref{fig:shape} shows the resulting shape of
the CD in comparison to the star--pulsar system as a function of $\eta_\infty\equiv
(L_\mathrm{sd}/c)/(\dot{M}v_\infty)$.

At the scales of interest, and given the very high speed of the stellar wind of
massive stars, the role of orbital motion in the shape of the CD is not expected
to be significant. The aberration or skew angle $\mu$ of the shock cap is
defined as the angle between the axis of symmetry of the shock cap and the line
of centers of the star and compact object. This angle is given by
$\tan\mu=v_\mathrm{orb}/v_\mathrm{w}$, where $v_\mathrm{orb}$ is the orbital
velocity and $v_\mathrm{w}$ is the speed of the slowest wind (i.e., the wind of
the star) at the contact surface. For the orbital parameters of \ls, for
example, the skew angle is below 23$^\circ$ at all moments along the orbit.

\section{Thermal X-ray emission}\label{sec:thermal}

\subsection{Cooling regime of the shocked stellar wind}
\label{sec:cooling}

On either side of the CD, the stellar and pulsar winds develop shocks. The
shocked pulsar wind is the candidate location for particle acceleration, which
would give rise to the observed broadband non-thermal emission
\citep{1997ApJ...477..439T}. On the other side of the CD, the stellar wind
develops a strong shock where the wind material is slowed down and heated.
The cooling efficiency of the gas determines the width of the cooling layer.
On one extreme, if radiative cooling is very efficient, the gas will be
rapidly cooled after the shock and will collapse into a thin dense layer on
the CD. The rapid cooling assures that all the wind kinetic energy flux
perpendicular to the shock is transformed into thermal emission. On the other
extreme, if radiative cooling is slow enough, the hot gas will flow along the
CD and cool adiabatically through expansion.  In the adiabatic regime, a
smaller fraction of the kinetic energy of the stellar wind is emitted in the
X-ray band. \cite{1992ApJ...386..265S} proposed the parameter
$\chi=t_\mathrm{rad}/t_\mathrm{esc}$ as a measure of the cooling regime of the
shocked material in colliding wind binaries. 
$\chi$ values significantly below (above) unity indicate that the shocked
stellar wind cools radiatively (adiabatically), whereas $\chi$ values around
unity indicate that a fraction of the incoming kinetic energy will be radiated
away and the rest will be dissipated through adiabatical expansion.


\subsection{Estimation of the expected X-Ray luminosity}
\label{sec:lumest}

Before we enter into the details of the calculation of the thermal X-ray
spectrum and luminosity, we will roughly derive the expected X-ray
luminosity as a function of the parameters of the system. As mentioned
above, we expect the X-ray luminosity to be a fraction $f$ of the wind
kinetic luminosity crossing the shock:
$L_\mathrm{X}=fL_\mathrm{kin}^\mathrm{sh}$. The value of $f$ is
strongly dependent on the value of $\chi$ at the shock, and as a first
approximation can be taken as $f\approx 1/(1+\chi)$ (see
Equation~(\ref{eq:kintoth})).
Additionally, the kinetic
luminosity deposited on the shock, $L_\mathrm{kin}^\mathrm{sh}$, will be
a fraction of the total kinetic luminosity of the stellar wind,
$L_\mathrm{kin}^\mathrm{tot}$, determined by the fraction of the total
solid angle of the stellar wind subtended by the shock. For a cone with
a half-opening angle of $\theta$, we obtain $L_\mathrm{X}\approx
f\frac{1}{2}(1-\cos\theta)L_\mathrm{kin}^\mathrm{tot}$, which can be
simplified up to $\mathcal{O}(\theta^4)$ as $L_\mathrm{X}\approx
f\frac{1}{4}\theta^2L_\mathrm{kin}^\mathrm{tot}$. 
To
obtain the estimate of the luminosity, we will here consider that the
wind reaches the shock front at terminal velocity. The angle $\theta$
can then be derived from Equation~(\ref{eq:difeq}) as $\pi\eta/(1+\eta)$,
which on a first-order approximation for small values of $\eta$ (i.e.,
dominance of the stellar wind) will be $\theta\approx\pi\eta$. Using
these approximations we obtain $L_\mathrm{X}\approx
f\frac{1}{4}\pi^2\eta^2L_\mathrm{kin}^\mathrm{tot}$.  As seen in
Equation~(\ref{eq:eta}), $\eta$ is proportional to $\lsd$.
Therefore, we expect the
emitted thermal X-ray luminosity from the shocked stellar wind to behave roughly
as:
\begin{equation}
    L_\mathrm{X}\approx1.2\times10^{32} 
    \left[\frac{\lsd}{10^{36}\,\ergs}\right]^2
    \left[ \frac{\dot{M}}{10^{-7}\,\msolyr}\right]^{-1}\ \ergs,
    \label{eq:lumest}
\end{equation}
where we have taken the wind terminal velocity as
$v_\infty=2400$\,km\,s$^{-1}$, and assumed that half of $L_\mathrm{kin}^\mathrm{sh}$
is converted into X-ray luminosity, i.e., $f=0.5$. This latter assumption is
consistent with a cooling regime between the adiabatic and radiative regimes,
i.e., $\chi\approx1$.

The above approximations will only hold for small values of $\eta$, i.e., for
the case of a dominant stellar wind. When the ram pressure of the pulsar wind is
similar to that of the stellar wind and $\eta$ approaches unity, the
half-opening angle of the CD will approach $\pi/2$. At this limit, the X-ray
luminosity will be even higher that for the previous case, and of the order of
\begin{equation}\label{eq:lumest2}
    L_\mathrm{X}\approx3\times10^{34} \left[ \frac{\dot{M}}{10^{-7}\,\msolyr}
    \right]\,\ergs.
\end{equation}

Gamma-ray binaries are non-thermal X-ray sources with luminosities of 
the order of several $10^{33}\,\ergs$. Therefore, as
Equations~(\ref{eq:lumest}) and (\ref{eq:lumest2}) show, 
pulsar spin-down luminosities moderately higher than $10^{36}\,\ergs$ 
would imply a significant thermal component in their X-ray spectrum. 

\subsection{X-Ray emission model details}\label{sec:modeldetails}


For a strong, steady, standing shock with a compression ratio
$\rho_0/\rho_\mathrm{w}=v_\mathrm{w\perp}/v_{0\perp}=4$ (where the subindices
$\mathrm{w}$ and $0$ indicate the properties of the gas immediately before and
after the shock, respectively), the gas will have an immediate postshock
temperature of 
\begin{equation}
    kT_0=\frac{3}{16}\mu m_p v_\mathrm{w\perp}^2 \approx 1.21 \left[\frac{\mu}{0.62}\right]
    \left[\frac{v_\mathrm{w\perp}}{1000\,\mathrm{km\,s^{-1}}}\right]^2\,\mathrm{keV,}
    \label{eq:t0}
\end{equation}
where $v_\mathrm{w\perp}$ indicates the preshock speed of the wind component
perpendicular to the shock front and $\mu$ is the average atomic weight. A value
of $\mu=0.62$ is appropriate for a fully ionized medium with solar abundances
\citep{1989GeCoA..53..197A}.

The X-ray emission of the shocked stellar gas is computed assuming that, in the
emitting plasma, conditions are equal to those at the immediate postshock
position, which is true as long as the gas remains subsonic.  This will provide
an estimate of how much of the kinetic energy of the incoming wind is radiated
away. Assuming the shocked gas acts as a calorimeter, in which all the incoming
perpendicular kinetic energy is either radiated away as X-ray emission or
dissipated through adiabatic expansion, the fraction of the kinetic energy
converted into X-ray emission will be
\begin{equation}
    L_\mathrm{X}=\frac{t_\mathrm{rad}^{-1}}{t_\mathrm{rad}^{-1}+t_\mathrm{esc}^{-1}}
    L_\mathrm{kin}^\mathrm{sh}, \label{eq:kintoth} 
\end{equation}
where $t_\mathrm{rad}$ is the radiative cooling timescale, $t_\mathrm{esc}$ is
the escape or flow dynamics timescale and $L_\mathrm{kin}^\mathrm{sh}$ is the
incoming kinetic energy. Given that $\dot{E}\propto t^{-1}$, the fraction of
times may be interpreted as the fraction of the radiated energy over the total
energy lost owing to radiation and escape.

For a gas cooling with a volumetric emission rate $\Lambda(T)$, the radiative
cooling timescale is given approximately by
\begin{equation}
    t_\mathrm{rad}=\frac{kT_0}{n_0\Lambda(T_0)},
    \label{eq:trad}
\end{equation}
where $n_0$ is the postshock number density, and in a strong shock will be four
times the preshock wind number density. Instead of using a pre-calculated
table for $\Lambda(T)$, it can be calculated using the \textsc{mekal} code used
to compute the emitted spectrum (see below).

The escape timescale is the time the heated gas takes to flow away from the
postshock location. Assuming that the stellar wind overpowers the pulsar wind,
we can use the distance from the
CD to the pulsar as the radius of curvature of the shock front, and thus as a
typical flow dynamics distance scale. The escape timescale can then be
approximated as $t_\mathrm{esc}=r_\mathrm{p}/v_0$. We assume that the wind
velocity component parallel to the shock is unaffected after traversing it, so
that the immediate postshock speed may be found as
$v_0^2=(v_\mathrm{w\perp}/4)^2+v_\mathrm{w\parallel}^2$.

We used the \textsc{mekal} code as implemented in the \textsc{spex} software
package \citep{1996uxsa.conf..411K} to generate a look-up table of emission
spectra for optically thin hot gas in collisional ionization equilibrium as a
function of the gas temperature $T$. 
We populated the look-up table with emission spectra in the 0.05--15\,keV range,
taking 250 logarithmically spaced temperatures from $10^{4}$ to $10^{9}$\,K.
Additionally, for each of the temperatures, the volumetric emissivity
rate $\Lambda(T)$ was computed by integrating the emission between photon
energies of 0.001 and 50\,keV.

In order to compute the total cumulative emission from the shocked stellar wind
we assumed that the X-ray emitter is infinitely thin and located on the CD
defined through Equation~(\ref{eq:difeq}). We defined a grid of 2000~bins in the
$y$-direction, away from the line of centers, and 30 azimuthal bins on the shock
front.  For each segment of the grid the immediate postshock temperature $T_0$
and X-ray luminosity was obtained through Eqs.~(\ref{eq:t0}) and
(\ref{eq:kintoth}).  Using the X-ray emission look-up table, we obtained the
intrinsic X-ray emission spectrum from each segment. Finally, we computed the
neutral hydrogen column density owing to the non-shocked stellar wind, from each
of the positions of the segments on the shock front to the observer.  We assumed
that neither the enhanced density of the shocked stellar wind nor the free
pulsar wind contribute significantly to the total column density. The column
density of the former can be estimated to be two orders of magnitude below the
wind column density \citep{2011MNRAS.411..193S}, and the latter is too rarefied
to absorb X-rays. We then computed the energy-dependent attenuation factor
corresponding to the found column density through the \texttt{wabs} subroutine
provided with Xspec~v12. In addition, the emission from those areal segments
eclipsed by the star was discarded. The absorbed emission from each of the areal
segments was added together to obtain the cumulative emitted spectrum of the
shocked stellar wind.

From the calculations described in this section, we can see what the general
characteristics of the shocked stellar wind emission will be. Along the line of
centers of the system we will find both the highest temperatures and the highest
densities owing to the perpendicular impact of the wind on the shock
front. The hardest X-rays will come from this location, but its volume is
relatively small and its luminosity may not dominate the final emission. Away
from this point temperatures will drop rapidly as the incoming wind will not be
perpendicular to the shock, whereas the volume increases, and density decreases.
This region will produce a softer X-ray spectrum. The final spectrum will be
a so-called multi-temperature X-ray spectrum, resulting from the emission of the
stratified temperature structure of the shocked gas.

\subsection{Caveats of the model}

The semi-analytic model detailed in the previous sections is robust given
that the assumptions made are fairly conservative. However, there are some
factors that affect the WCR that cannot be accounted for in
our semi-analytic framework. Detailed hydrodynamical simulations would be
required to take into account effects such as the following.
\begin{itemize}
    \item In the radiative cooling regime the CD may be disrupted by
        thin-shell instabilities
        \citep{1992ApJ...386..265S,2009MNRAS.396.1743P}. Additionally,
        stellar wind clumping would likely disturb the shocked wind region to
        some extent. If this perturbation is propagated to the shocked pulsar
        wind, it would have implications for both the thermal and non-thermal
        emission.
    \item We modeled the radiative output of the postshock gas through the
        ratio of radiative to total cooling timescales in
        Equation~(\ref{eq:kintoth}), and assumed an infinitely thin cooling layer.
        Both the intrinsic X-ray luminosity and cooling layer width could be
        calculated self-consistently through hydrodynamical simulations.
        Whereas the results are likely to be very similar in the radiative
        cooling regime limit, they will probably differ in the adiabatic limit.
        In the adiabatic limit of the postshock flow, the flow may become
        supersonic at a distance smaller than $r_\mathrm{p}$, cooling the gas
        and reducing the radiative outcome. This effect would not be important
        at scales close to the apex, but could affect the contribution produced
        at the periphery of the system.
\end{itemize}

\subsection{Method of spin-down luminosity upper limit derivation}
\label{sec:ulimderivation}

To accurately compare the output of the model
described in Sections~\ref{sec:dynamic}
and \ref{sec:thermal} with the observed X-ray spectra of gamma-ray binaries,
the computed X-ray spectra can be formatted as a FITS table model in order to
load it into the spectral analysis software package \emph{Sherpa}
\citep{2001SPIE.4477...76F}. The FITS table model is generated from the
cumulative spectra for 25 pulsar spin-down luminosities spaced logarithmically
between $10^{36}$\,\ergs and $5\times10^{37}$\,\ergs, taking a constant value of
the mass-loss rate and orbital inclination.

The X-ray spectra observed at a certain orbital phase can be assigned a source
model corresponding to an interstellar medium neutral hydrogen absorption model
(\emph{A}), affecting both the non-thermal power law (\emph{P}) and the computed
shocked stellar wind thermal emission (\emph{Th}):
$A_\mathrm{ISM}\times(P+\mathit{Th})$.
%
%
The goal is then to find for which value of the spin-down luminosity the thermal
component begins to affect the shape of the spectrum and the fit worsens
significantly. We used the \texttt{confidence} \emph{Sherpa} command to evaluate
the 99.7\% ($3\sigma$) confidence level range for acceptable values of the
spin-down luminosity. If no thermal features are present in the observed
spectrum, the lower bound will be consistent with zero, i.e., lack of thermal
component. The upper bound, on the other hand, provides us with the upper limit
to the spin-down luminosity of the pulsar.

\section{Application of the model to \ls}\label{sec:appls}

The system \ls\ is located at $2.5\pm0.1$\,kpc and contains a compact object
with a mass between $1.4$ and $5\,\msol$, orbiting an O6.5V((f)) donor star
every $3.90603\pm0.00017$~days \citep{2005MNRAS.364..899C} in a mildly eccentric
orbit \citep[$e=0.24\pm0.08$;][]{2011MNRAS.411.1293S}. The detection of
elongated asymmetric emission in high-resolution radio images was interpreted as
mildly relativistic ejections from a microquasar jet and prompted its
identification with an EGRET HE gamma-ray source
\citep{2000Sci...288.2340P,2002A&A...393L..99P}. However, recent Very Long
Baseline Array observations by \cite{2008A&A...481...17R} show morphological
changes on short timescales that might be consistent with a pulsar binary
scenario. It shows variable periodic emission in the X-ray, and HE and VHE
gamma-ray bands. The X-ray lightcurve has an orbital modulation
\citep{2009ApJ...697..592T}, superposed by short timescale features that are
quite stable over the years \citep{2009ApJ...697L...1K}. The maximum and minimum
emission phases in the periodic X-ray lightcurve are located around the inferior
and superior conjunctions, so a geometrical effect for the modulation has been
suggested. However, inferior and superior conjunctions are located close to
apastron and periastron, and the broadness of the peaks, well fitted by a
sinusoidal function, makes it difficult to discern between a physical or
geometric origin of the modulation. Whereas the VHE gamma-ray lightcurve has a
modulation similar to the X-ray one \citep{2006A&A...460..743A}, the HE
gamma-ray flux is lower at the inferior conjunction ($0.45<\phi<0.9$) and higher
at superior conjunction ($0.9<\phi<0.45$). In addition, at superior conjunction
it shows a spectral cutoff at $E=1.9\pm0.5$~GeV, not statistically required for
inferior conjunction data \citep{2009ApJ...706L..56A}. 

The radial and powerful wind of \ls\ makes it an ideal system to apply the model
presented in this work. The shocked gas cooling regime (see
Section~\ref{sec:cooling}) will be closer to the radiative than to the adiabatic
limit, with $\chi$ values around unity. Therefore, it seems safe to apply the
thermal emission model presented in Section~\ref{sec:modeldetails} to \ls.

\subsection{Stellar mass-loss rate and pulsar spin-down luminosity in \ls}
\label{sec:lsprop}

\begin{figure}
    \includegraphics[width=\figwidth]{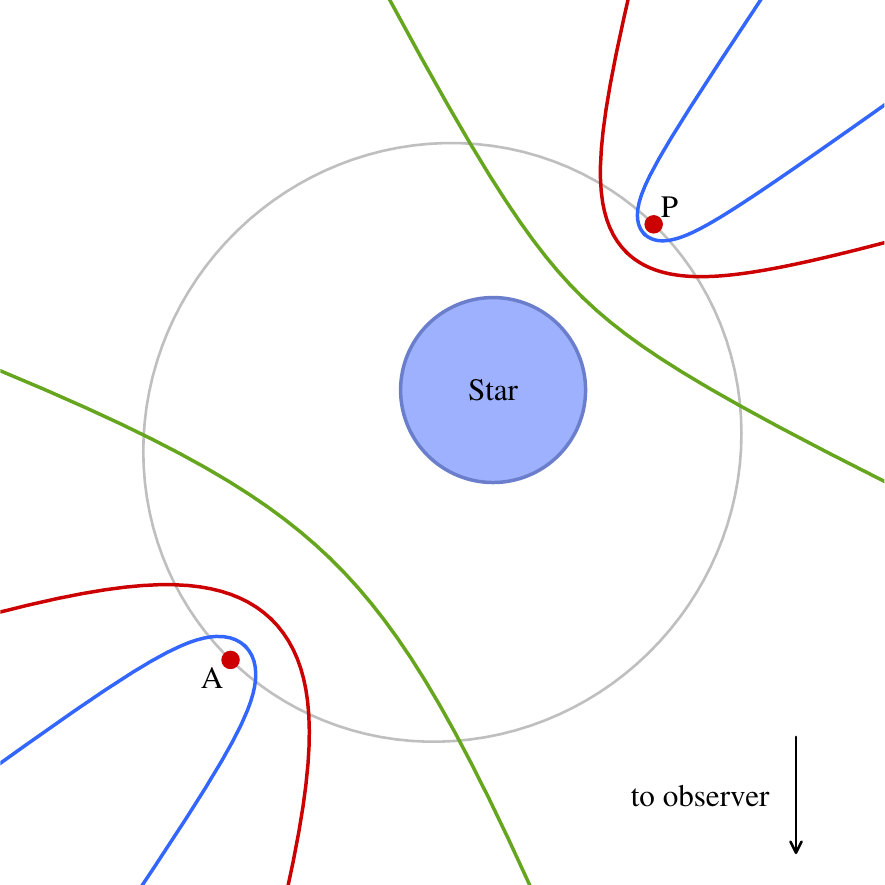}
    \caption{Sketch of the orbit of the compact object around the companion star
    in \ls. The star (but not the pulsar) is at scale. The two marked positions
    correspond to periastron (P) and apastron (A). For each of these positions
    the shape of the CD is shown for $\eta_\infty=0.0025,0.025$ and $0.25$
    (blue, red, and green lines, respectively).}
    \label{fig:orbit}
\end{figure}

As explained in Section~\ref{sec:dynamic}, the shape of the CD depends only on the
parameter $\eta_\infty$, which in turn depends only on the ram pressure of the
winds on either side of the CD. Figure~\ref{fig:orbit} shows the shape of the CD
in comparison to the orbit of \ls\ at periastron and apastron for different
values of $\eta_\infty$. \cite{2005MNRAS.364..899C} derived that the O6.5V((f))
star in \ls\ has a stellar radius of $R_\star=9.3\,\mathrm{R}_\odot$, its wind
has a terminal velocity of $v_\infty=2400$\,km\,s$^{-1}$ and the orbital inclination is
in the range 20$^\circ$--70$^\circ$. Therefore, the only parameters still
unknown that affect the shape of the CD, and thus the thermal X-ray emission,
are the stellar mass-loss rate, $\dot{M}$, and the pulsar spin-down luminosity,
$\lsd$. 

\cite{2011MNRAS.411.1293S} performed an optical photometric and spectroscopic
campaign on \ls, and determined the mass-loss rate of the star to be $3.7$ to
$4.8\times10^{-7}\,\msolyr$ from H$\alpha$ line measurements. However,
H$\alpha$ line fitting is a diagnostic known to be affected by wind clumping,
and owing to its $\rho^2$ dependence could easily overpredict mass-loss rates by
a factor $\sim$2 \citep{2004A&A...413..693M}. \cite{2010ASPC..422...77B} studied
the effect of wind clumping in the absorption of X-ray emission from \ls, but it
must be noted that \cite{2011MNRAS.411.1293S} did not detect any evidence of
dense wind clumps in the analysis of stellar wind emission lines.
The lack of
variability of the column density measured in X-ray observations along the orbit
may also be used to constrain the maximum value of mass-loss rate.
\cite{2007A&A...473..545B} concluded that, considering a non-thermal X-ray
emitter inside the binary system, the mass-loss rate could not exceed a few
$10^{-8}$\,\msolyr.
Adopting a model in which the non-thermal emission is extended, as
expected in the PWS scenario, this upper limit is relaxed up to
$1.5\times10^{-7}\,\msolyr$ \citep{2011MNRAS.411..193S}. 
In the following calculations we considered the whole range of mass-loss rates
discussed here, i.e., from $5\times10^{-8}\,\msolyr$ to
$4.8\times10^{-7}\,\msolyr$. For some cases, in which a single representative
mass-loss rate is used to test other properties of the thermal emission, we
chose the midpoint of the range: $\dot{M}=2.65\times10^{-7}\,\msolyr$.


The spin-down luminosity of the putative pulsar in \ls\ is not known as it
has never been measured directly. A lower limit may be derived from efficiency
considerations taking into account the HE gamma-ray emission detected by
\emph{Fermi}/LAT \citep{2009ApJ...706L..56A}. If the HE gamma-ray emission has a
magnetospheric origin, the detected flux above 100\,MeV of
$2.55\times10^{-10}$\,\ergcms\ implies a spin-down luminosity between 0.2 and
$30\times10^{36}$\,\ergs, depending on the beaming model, with higher
values favored by the high energy spectral cutoff \citep[see][for an
explanation of the derivation of these values for the case of
LS\,I\,+61\,303]{2011A&A...527A...9Z}. However, the observed orbital
modulation of the HE gamma-ray emission, and the lack of pulsations,
indicate that its origin might not be magnetospheric but rather IC. 

The minimum non-thermal luminosity required for the \emph{Fermi} emission of
LS~5039 can also be estimated in the context of IC radiation. To obtain a
conservative estimate, one can assume that electrons are injected only in the
relevant energy range, $\sim$1--100\,GeV, their velocities are isotropically
distributed, and have time to
cool down only through IC. Stellar IC leading to photons in the \emph{Fermi} band
occurs in the Thomson regime. Since this process is strongly anisotropic
($L_\mathrm{IC}\propto (1-\cos\theta_\mathrm{IC})^{(p+1)/2}$ with $p$ being the
electron power-law index; e.g., \citealt{1992A&A...256L..27D}), one needs to
account for the IC interaction angle ($\theta_\mathrm{IC}$) between the observer
line of sight and the seed photon direction. Under these
conditions, the injection luminosity of electrons can be written as 
\begin{equation}
L_\mathrm{e}\approx
1.7\,\frac{2^{\alpha+1}\,L_\mathrm{GeV}}{2(\alpha+1)(1-\cos\theta_\mathrm{IC})^{\alpha}}\,, 
\end{equation} 
where the factor 1.7 accounts for the luminosity radiated below 100\,MeV,
$\alpha=(p+1)/2$, and $\theta_\mathrm{IC}\le\pi/2$ for phases around the
inferior conjunction. Taking $L_\mathrm{GeV}\approx 2\times 10^{35}\,\ergs$
during those orbital phases \citep{2009ApJ...706L..56A}, $p\sim 3$ to explain
the {\it Fermi} spectrum, and a rather conservative $\theta_\mathrm{IC}\approx
60^\circ$, one obtains $L_\mathrm{e}\approx 2\times 10^{36}\,\ergs$. However,
escape, adiabatic, and perhaps synchrotron losses could easily increase
$L_\mathrm{e}$ by a factor of several. In addition, not all the $L_\mathrm{sd}$
will go to non-thermal particles\footnote{Not only may the acceleration
efficiency be smaller than 100\%, but for higher spin-down luminosities the
opening angle of the CD will increase and only around half of the pulsar wind
luminosity will be shocked.}, and the injection electron distribution is
expected to be broader than $\sim$1--100\,GeV. From all this, $L_{\rm sd}\ga
2\times 10^{37}\,\ergs$ seems more realistic. Doppler boosting has not been
accounted for but, although it can reduce the required energy budget in certain
orbital phases, it will increase it in others. Note that for
$\theta_\mathrm{IC}\approx 90^\circ$ and phases between inferior and superior
conjunction, i.e., with the flow bulk motion unlikely pointing to us, still
$L_\mathrm{e}\ga 10^{36}\,\ergs$, and thus $L_{\rm sd}\ga 10^{37}\,\ergs$. The
GeV emitter may be extended, thus relaxing the $\theta_\mathrm{IC}$ dependence,
but it should not change significantly our conclusion. \label{sec:doppler}

\subsection{Comparison to X-Ray data}\label{sec:xray}

\ls\ has been observed at different positions along its orbit by many X-ray
observatories \citep[see][for a summary]{2008IJMPD..17.1867Z}. Recently, a
very long observation with \emph{Suzaku} provided an uninterrupted lightcurve of
the source during an orbit and a half \citep{2009ApJ...697..592T}, and a
comparison with previous observations indicated that it was quite stable
over periods of up to a decade \citep{2009ApJ...697L...1K}. The
\emph{Suzaku} observation provides a very good orbital coverage, but we have
chosen, to compare the output of our numerical model, two \emph{XMM-Newton}
observations performed in 2005 \citep{2007A&A...473..545B}. The
superior effective area of \emph{XMM-Newton} gives us the possibility to
extract very good soft X-ray spectra in relatively short observations of
duration $\Delta\phi\sim0.04$. These short
observations assure that the ambient conditions do not
vary significantly and the measured spectra can be accurately compared to
spectra calculated for a given phase.
These two $\sim$15\,ks observations were performed at periastron
(\dataset[ADS/Sa.XMM#0202950301]{ObsId 0202950301}, orbital phases 0.02--0.05) and
apastron (\dataset[ADS/Sa.XMM#0202950201]{ObsId 0202950201}, orbital
phases 0.49--0.53) during the same orbital period, thus probing the two extreme
separations between the star and the compact object.

For each of the two observations we processed the data through the standard
procedure with version 10.0 of the \emph{XMM-Newton} Science Analysis Software.
We filtered out the periods of high particle background for each of the three
detectors aboard \emph{XMM-Newton}, and extracted source and background spectra
from the clean event files, as well as the corresponding redistribution matrix
and ancillary response files. We binned the extracted source spectra to a
minimum of 20 counts per bin, avoiding an oversampling of the detector energy
resolution by more than a factor three. 

Before introducing our model of the thermal X-ray emission of the shocked
stellar wind, we performed a fit of the data to an absorbed power law using the
spectral analysis package \emph{Sherpa} bundled with CIAO~4.3. The fit results
for this strictly non-thermal source model may be found in the third column of
Table~\ref{tab:fitresults}.

We assigned each of the two groups of spectra (periastron (p) and apastron (a))
the source model described in Sections~\ref{sec:ulimderivation}:
$A_\mathrm{ISM}\times(P^\mathrm{p,a}+\mbox{\emph{Th}}^\mathrm{p,a})$.
Whereas the normalization and photon index parameters of the power-law component
were left free for the spectral fitting, the interstellar absorption and the
spin-down luminosity of the pulsar were kept constant between the periastron and
apastron source models. We note that such a choice for the source model implies
the assumption that the non-thermal emission is not significantly affected by
circumstellar absorption, whereas it affects the thermal emission and is already
computed in the component \emph{Th}. This would happen for either extended
non-thermal emitters or low circumstellar absorption (see
Section~\ref{sec:discmdot} below for a discussion).

As expected from the previous estimate (see Section~\ref{sec:lumest}), the X-ray
luminosity of the thermal component increased monotonically with the value of
the spin-down luminosity of the pulsar. In order to derive the spin-down
luminosity upper limit, we performed a simultaneous fit of the six spectra (three
detectors for each of the two orbital phases). In all cases, the best fit
corresponded to the lowest level of emission from the thermal component, i.e.,
the non-thermal component alone provided the best fit to the data. The upper
bound, as explained in Sec.~\ref{sec:ulimderivation}, corresponds to the upper
limit to the pulsar spin-down luminosity.

\begin{deluxetable}{llcc}
\tablecaption{Results of the X-Ray Spectral Analysis\label{tab:fitresults}}
\tablewidth{0pt}
\tablehead{
Orb.~Phase & Parameter                        & \multicolumn{2}{c}{Source Model} \\\cmidrule{3-4}
           &                                        & $A_\mathrm{ISM}\times P$          & $A_\mathrm{ISM}\times(P+\mbox{\emph{Th}})$ }
\startdata
           & $\Gamma$                               & $1.57\pm0.04$                     & $1.56\pm0.03$                      \\
Periastron & $F(P)$\tablenotemark{a}                & $\phantom{1}7.1\pm0.6\phantom{1}$ & $\phantom{1}6.6\pm0.4\phantom{1}$  \\
           & $F(\mbox{\emph{Th}})$\tablenotemark{a} & \nodata                        & $0.37$                             \\\midrule
           & $\Gamma$                               & $1.47\pm0.03$                     & $1.42\pm0.03$                      \\
Apastron   & $F(P)$\tablenotemark{a}                & $11.4\pm 0.7\phantom{1}$          & $10.9\pm0.6\phantom{1}$            \\
           & $F(\mbox{\emph{Th}})$\tablenotemark{a} & \nodata                        & $0.58$                             \\\midrule
           & $N_\mathrm{H}$\tablenotemark{b}        & $0.67\pm0.02$                     & $0.69\pm0.02$                      \\
           & $\lsd$                                 & \nodata
           & $4.09\times10^{36}\,\ergs$         \\
           & $\chi^2$/dof                           & 467.5/612                         & 477.1/612                          
\enddata
\tablenotetext{a}{Unabsorbed X-ray fluxes of the
non-thermal power law (\emph{P}) and the thermal shocked stellar wind
(\emph{Th}) components in the range 0.3--10\,keV are given in units of
$10^{-12}\,\ergcms$.} 
\tablenotetext{b}{Interstellar column density is in units of
$10^{22}$\,cm$^{-2}$.}
\end{deluxetable}

\section{Results}\label{sec:results}

\subsection{Thermal X-Ray spectra}

{
We used the dynamical, radiative, and absorption models described in
Sections.~\ref{sec:dynamic} and \ref{sec:thermal} to calculate the thermal
X-ray spectra emitted by the shocked stellar wind of \ls. 
 For the properties of this system we find that the cooling regime of
the shocked gas is an intermediate regime, with $\chi$ around unity. As
an example, for $\dot{M}=2.65\times10^{-7}\,\msolyr$ and
$\lsd=3\times10^{36}\,\ergs$ ($\eta_\infty=0.025$), we obtain radiative
cooling and escape timescales at the CD apex during periastron of
$9.1\times10^3$\,s and $7.7\times10^3$\,s, respectively, resulting in
$\chi\simeq1.18$. The maximum postshock temperature reached is
$T_0=2.43$\,keV at the apex, with a postshock number density of
$n_0\simeq2\times10^{10}$\,cm$^{-3}$.}

\begin{figure}
    \includegraphics[width=\figwidth]{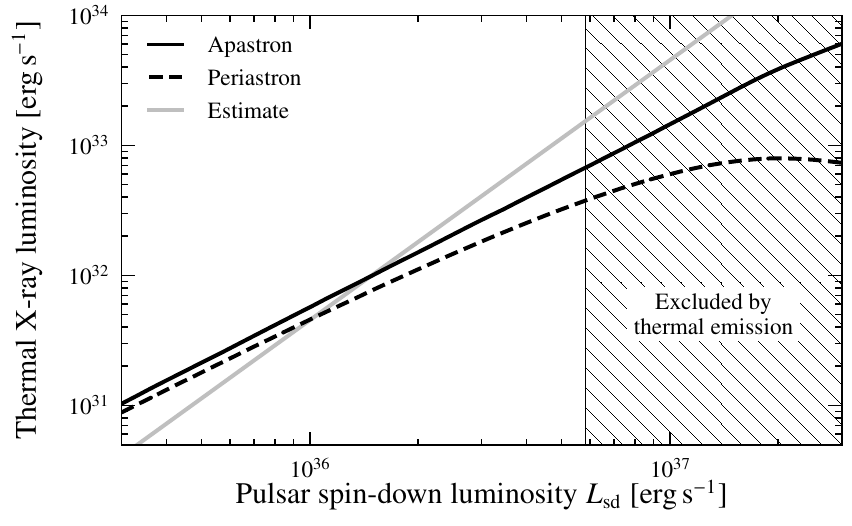}
    \caption{Thermal emission luminosity in the 0.3--10\,keV range as a function
    of the spin-down luminosity of the pulsar. The fluxes at periastron (solid)
    and apastron (dashed) are shown. In addition, the rough estimate of
    Equation~(\ref{eq:lumest}) is shown as a gray line. The range of pulsar
    spin-down luminosities excluded by the thermal emission is shown as a
    hatched region. In all cases, a mass-loss rate of
    $2.65\times10^{-7}\,\msolyr$ was assumed.}
    \label{fig:thfluxes}
\end{figure}

In accordance with the estimates of Section~\ref{sec:lumest}, we found
that in general the computed thermal X-ray luminosities increase
monotonically with the spin-down luminosity of the pulsar. In
Figure~\ref{fig:thfluxes} we show the 0.3--10\,keV X-ray luminosity of the
thermal emission from the shocked stellar wind as a function of the
pulsar spin-down luminosity, taking
$\dot{M}=2.65\times10^{-7}\,\msolyr$. For scenarios with
$\eta_\infty\ll1$ the behavior is generally in good agreement with
Equation~(\ref{eq:lumest}), and the thermal X-ray luminosity can be
acceptably explained by a function of the form
$L_\mathrm{X}\propto\lsd^\alpha$ with $\alpha={1.381\pm0.004}$.  This
value is significantly lower than that predicted by the estimate of
Equation~(\ref{eq:lumest}). This could be related to the lack of correction
for the obliquity of the incoming flow with respect to the shock in the
estimate, which could result in an underestimation for lower
$\eta_\infty$ and an overestimation at higher $\eta_\infty$, where a
larger fraction of the incoming wind impacts obliquely.  At higher
$\eta_\infty$, the proximity of the CD to the stellar surface at
periastron provokes a decrease in the velocity of the wind crossing the
shock, and therefore a decrease in the total X-ray luminosity. 

\begin{figure}
    \includegraphics[width=\figwidth]{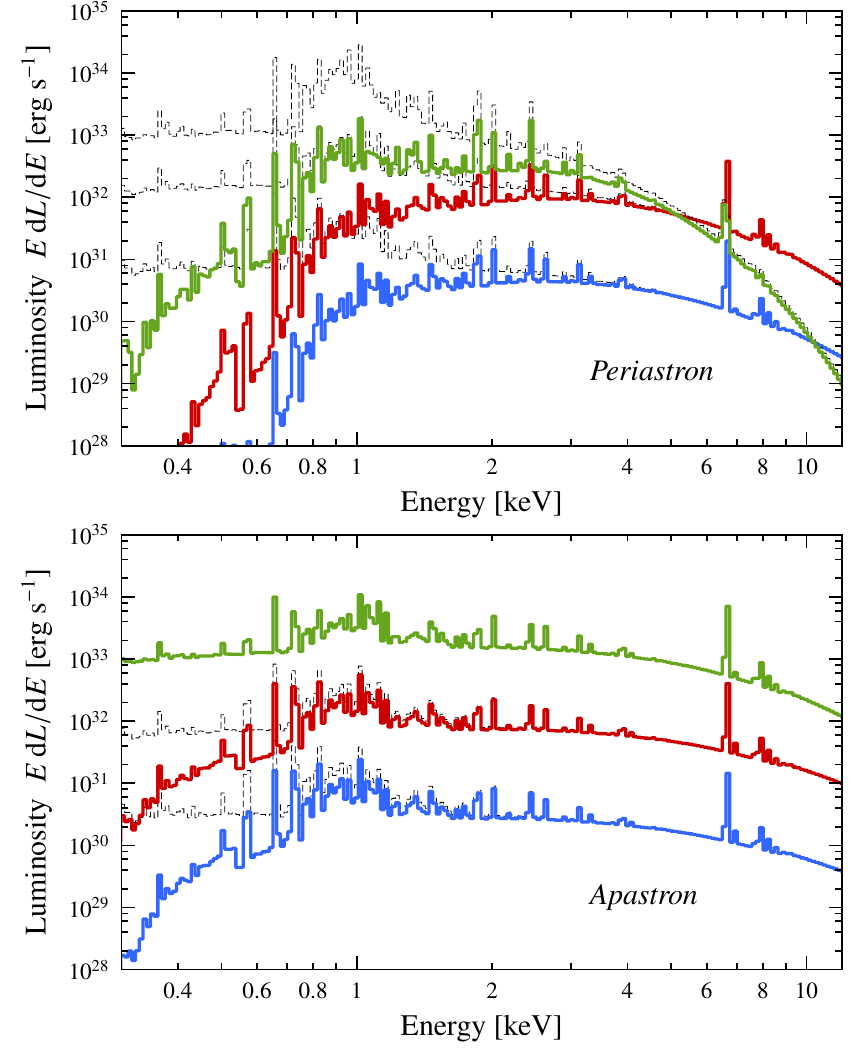}
    \caption{X-ray emission computed from the shocked stellar wind of \ls\ at
    periastron (top) and apastron (bottom), taking
    $\dot{M}=2.65\times10^{-7}\,\msolyr$. The different spectra for each orbital
    phase correspond to different values of pulsar spin-down luminosity $\lsd$:
    0.3 (blue, bottom), 3 (red, middle), and 30 (green, top)
    times $10^{36}\,\ergs$, corresponding to $\eta_\infty=0.0025$, 0.025, and
    0.25, respectively. The thin dashed lines indicate the intrinsic spectra,
    whereas the thick solid ones are the ones absorbed owing to the stellar
    wind.} \label{fig:specresults}
\end{figure}

Figure~\ref{fig:specresults} shows the obtained intrinsic and absorbed
spectra for different values of the spin-down luminosity from
$3\times10^{35}\,\ergs$ to $3\times10^{37}\,\ergs$. The mass-loss rate
of $2.65\times10^{-7}\,\msolyr$ assumed for these spectra provides a
dense matter field in the binary. Its effect can be seen as a very
strong absorption at energies below 1\,keV for the periastron spectrum.
The column density at this phase is of the order of
$10^{22}$\,cm$^{-2}$, even higher that the measured interstellar one. In
contrast, the apastron spectrum for the highest spin-down luminosity
shows negligible circumstellar absorption. At apastron, the pulsar would
be close to inferior conjunction, and the high opening angle of the CD
would mean that most of the thermal X-ray emission only has to travel
through the unshocked pulsar wind to reach the observer (see
Figure~\ref{fig:orbit}), thus being barely absorbed. As discussed in
Section~\ref{sec:modeldetails}, the hardest X-rays will come from the
region around the apex of the CD. For high values of the spin-down
luminosity, the apex of the CD at periastron will be close to the
stellar surface, and the velocity of the incoming wind will be low owing
to the wind $\beta$-velocity law. Therefore, the X-ray emission from the
apex will be softer, an effect that can be seen in the high energy part
of the topmost spectrum of Figure~\ref{fig:specresults}. For the
parameters used to compute this spectrum, the stagnation point of the CD
is at only 0.39 stellar radii from the surface of the star, and the wind
velocity at the shock is about one third of the terminal velocity. Note
how this effect is not apparent in the corresponding apastron spectra,
for which the stagnation point is much farther from the stellar surface.

\subsection{Spin-down luminosity upper limits}

\begin{figure}
    \includegraphics[width=\figwidth]{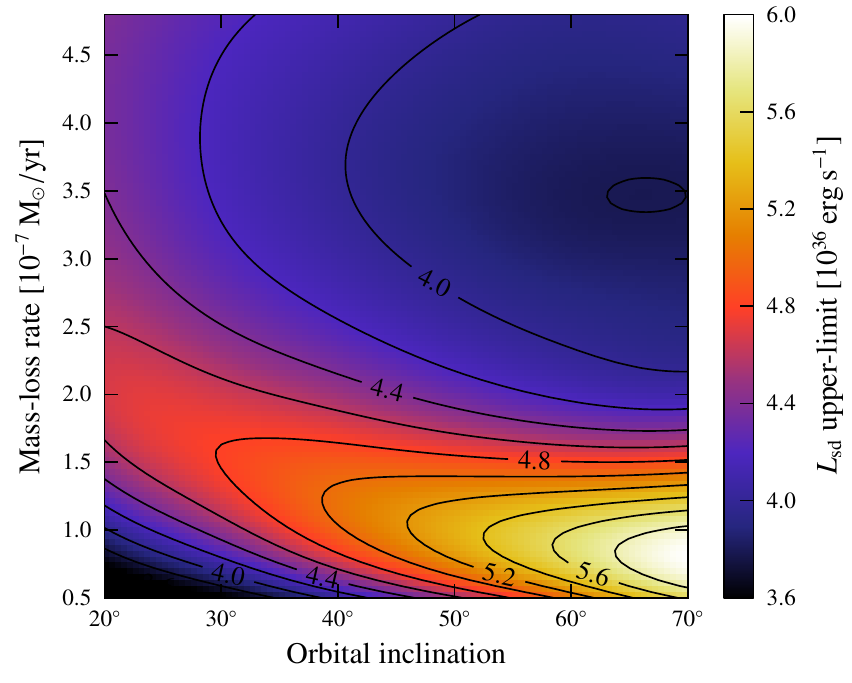}
    \caption{Spin-down luminosity upper limits as a function of orbital
    inclination and stellar mass-loss rate.
    \label{fig:ulims_imdot}}
\end{figure}

To derive the upper limits of the pulsar spin-down luminosity, we chose to
explore the stellar mass-loss rate values mentioned in Section~\ref{sec:lsprop},
corresponding to the estimates of \cite{2007A&A...473..545B},
\cite{2011MNRAS.411..193S}, and \cite{2011MNRAS.411.1293S}:
$5\times10^{-8}\,\msolyr$, $1.5\times10^{-7}\,\msolyr$, and
$4.25\times10^{-7}\,\msolyr$, respectively. The orbital inclination
($i=20^\circ\mbox{--}70^\circ$; \citealt{2005MNRAS.364..899C}) will
affect the importance of circumstellar absorption at periastron in the
emitted spectrum. The result of the pulsar spin-down luminosity
upper limit determination, for the whole ranges of mass-loss rate and orbital
inclination, is shown in Figure~\ref{fig:ulims_imdot}. A summary of the results
for the extremes and midpoints of the ranges can be seen in
Table~\ref{tab:ulims}.
%

\begin{deluxetable}{lcc}
    \tablecaption{Pulsar Spin-down Luminosity Upper Limits
    \label{tab:ulims}}
    \tablewidth{0pt}
\tablehead{
Mass-loss Rate & {Inclination} & $L_\mathrm{sd}$ Upper Limit \\
(\msolyr) &  &  ($10^{36}\,\ergs$) 
}
\startdata
                             & 20$^\circ$ & 3.3 \\
$5\times10^{-8}$ & 45$^\circ$ & 4.3 \\
                             & 70$^\circ$ & 5.6 \\\midrule
                             & 20$^\circ$ & 4.5 \\
$1.5\times10^{-7}$       & 45$^\circ$ & 4.9 \\
                             & 70$^\circ$ & 4.8 \\\midrule
                             & 20$^\circ$ & 4.4 \\
$4.25\times10^{-7}$          & 45$^\circ$ & 4.0 \\
                             & 70$^\circ$ & 3.9
\enddata
\end{deluxetable}

\begin{figure} 
    \includegraphics[width=\figwidth]{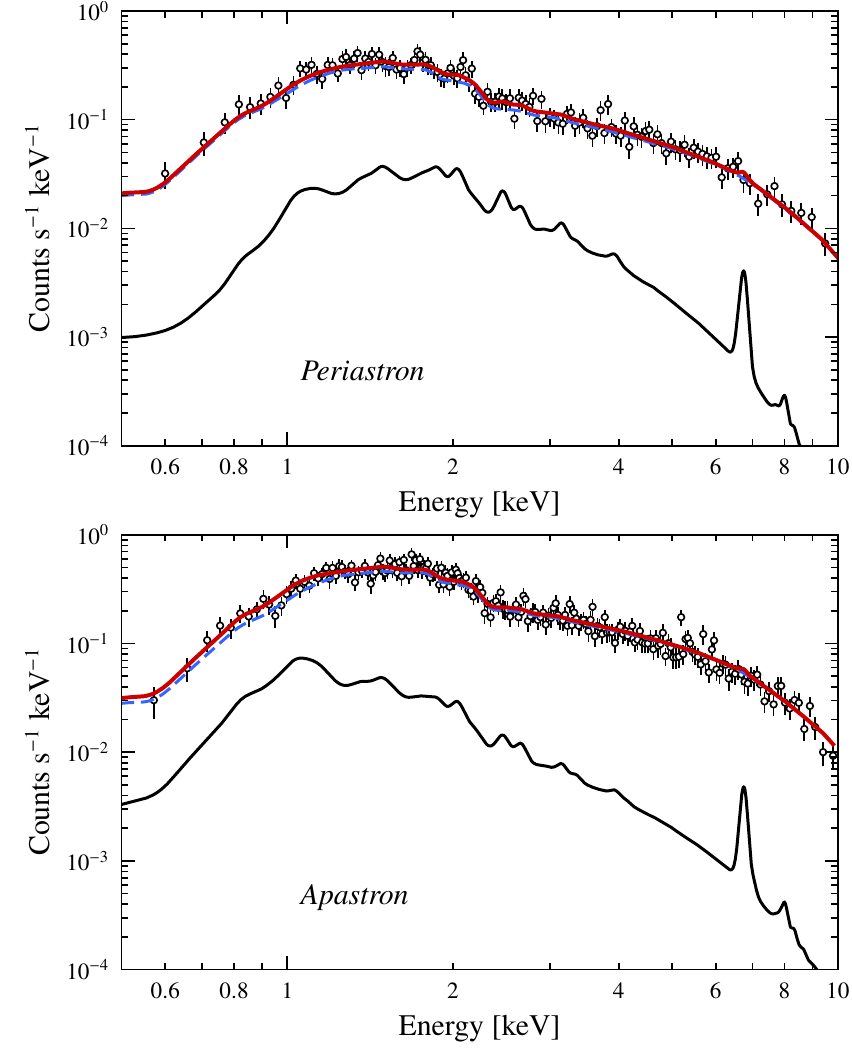} 
    \caption{X-ray spectra of \ls\ at periastron (top) and apastron
    (bottom).  The open circles are the data from the pn detector. The
    MOS data are fit simultaneously but not shown here for clarity. The
    lines indicate the best-fit spectral model for the spin-down
    luminosity fixed at the upper limit for moderate values of orbital
    inclination ($i=45^\circ$) and stellar mass-loss rate
    ($\dot{M}=2.65\times10^{-7}\,\msolyr$):
    $\lsd=4.09\times10^{36}\,\ergs$. The thick solid line is the sum of
    the non-thermal power-law component (thin dashed) and the shocked
    stellar wind thermal component (thin solid). The energy resolution
    of the thermal component is lower than in
    Figure~\ref{fig:specresults}, which results from folding the spectral
    models with the energy redistribution matrix of the detector.
    \label{fig:specfit}} 
\end{figure}

We found that pulsar spin-down luminosities above $6\times10^{36}\,\ergs$
are excluded by the lack of spectral thermal features in the observed periastron
and apastron spectra, for any combination of mass-loss rate and orbital
inclination. This upper limit corresponds to the extreme inclination of
$70^\circ$, and for moderate orbital inclinations the upper limit is at all
times below $5.5\times10^{36}\,\ergs$. Although the wind kinetic luminosity
available for $\dot{M}=5\times10^{-8}\,\msolyr$ is lower than for the other
cases, the upper limits derived are at least as strict as for higher mass-loss
rates. We found that the reduced absorption owing to the lower density of the
stellar wind results in brighter emission below 1\,keV. In the last column of
Table~\ref{tab:fitresults}, we show the best-fit results of the thermal plus
non-thermal source model when fixing the spin-down luminosity at
the upper limit we found. The thermal component was computed taking the
midpoint values of mass-loss rate and orbital inclination.
As compared to the non-thermal source model, the best-fit
power law is slightly harder and its unabsorbed flux is $\sim$5\% lower.
An illustration of the thermal and non-thermal contributions to the
observed spectrum is shown in Figure~\ref{fig:specfit}.

\section{Discussion}\label{sec:discussion} 

\subsection{On the stellar mass-loss rate}\label{sec:discmdot}

The stellar mass-loss rate plays a double role in the PWS scenario.
On one hand it determines the shape of the CD and the kinetic luminosity
that is shocked onto the CD. On the other hand, it determines the importance of
photoelectric absorption in the X-ray spectra of the thermal and non-thermal
radiation sources. 

The lack of X-ray absorption variability in \ls\ along the orbit
indicates that, if the non-thermal emitter is inside the orbital system,
the stellar wind density must be low. However, a very low value will
result in the pulsar wind overpowering the stellar wind for reasonable
values of the pulsar spin-down luminosity.  The maximum mass-loss rate
compatible with a variation of the neutral hydrogen column density along
the orbit below $10^{21}$\,cm$^{-2}$ \citep{2007A&A...473..545B} is of
$\sim5\times10^{-8}\,\msolyr$. For such a low mass-loss rate, the pulsar
wind could impact on the stellar surface at periastron for reasonable
values of spin-down luminosity. However, the lack of orbital variations
of the UV spectra of \ls\ \citep{2004ApJ...600..927M} indicate that the
stellar surface is not perturbed at any point along the orbit.  This
places an upper limit on the spin-down luminosity of
$\sim7\times10^{36}\,\ergs$.
However, if the star has a strong enough surface magnetic field, then
magnetic pressure could be enough to balance the pulsar wind away from
the surface \citep{1990ApJ...358..561H}. For a surface magnetic field of
$B_\star=50$\,G and assuming a dipole field, the spin-down luminosity
could be as high as $3\times10^{37}\,\ergs$ without the CD collapsing
onto the stellar surface.  Note, however, that
$\dot{M}=5\times10^{-8}\,\msolyr$ and $B_\star=50$\,G imply a wind
magnetic to kinetic energy ratio close to unity.

If a significant fraction of the non-thermal emitter is not located deep
inside the binary system, the photoelectric absorption of the thermal
and non-thermal X-ray emission could then be decoupled (as we have done
in Section~\ref{sec:ulimderivation}), since their locations would be
different. For this scenario, the stellar mass-loss rate could be as
high as a few times $10^{-7}\,\msolyr$, as measured by
\cite{2011MNRAS.411.1293S}.  Additionally, such a location would favor
the acceleration of particles up to the very high energies required to
account for the VHE gamma-ray spectrum, as well as avoid strong pair
creation absorption of VHE gamma rays
\citep{2008MNRAS.383..467K,2008A&A...489L..21B}. 

\subsection{On the pulsar spin-down luminosity}

The found upper limits of the pulsar spin-down luminosity pose strong
constraints on further modeling of the PWS scenario for non-thermal emission. As
discussed in Section~\ref{sec:lsprop}, the pulsar spin-down luminosity should be
at least a few times $10^{36}\,\ergs$ in order to account for the GeV emission.
The upper limit we have found from the thermal X-ray emission
($\lsd<6\times10^{36}\,\ergs$) indicates that the spin-down luminosity must be
very close to this value in order for the PWS scenario to be consistent. A
detailed modeling of the GeV emission from \ls\ is required to accurately check
the consistency of the observed GeV luminosity and the PWS scenario.
Unfortunately, the structure and location of the GeV emitter is not well known,
and IC emission is very sensitive to the $\theta_\mathrm{IC}$-value, which makes
difficult to obtain conclusive results.  Nevertheless, the lower limit derived
here for $L_\mathrm{e}$, in the IC scenario, is still robust enough to show
that, likely, $L_\mathrm{e}/L_{\rm sd}\rightarrow 1$. It might appear that such
a strong constraint can only be achieved at very low $\eta_\infty$, but the
effect of the orbital motion on the large-scale structure allows for the pulsar
wind to be terminated even in the direction opposite to the star up to
$\eta_\infty\sim1$ \citep{2011arXiv1105.6236B}. We note that the constraint on
the efficiency is slightly less restrictive for high orbital inclinations
($i\gtrsim45^\circ$) and low stellar mass-loss rates
($\dot{M}\lesssim2\times10^{-7}\,\msolyr$).

The recent detection of \psr\ at GeV energies \citep{2011ApJ...736L..10T,
2011ApJ...736L..11A} provides a perfect comparison of the efficiency of
spin-down luminosity to non-thermal emission conversion efficiency. During its
highest state of emission the total gamma-ray luminosity was comparable to the
pulsar spin-down luminosity, which combined with the recent reclassification of
the optical star \citep{2011ApJ...732L..11N} results in nearly unbearable energy
requirements \citep{2011arXiv1104.0211K}.  Even considering Doppler boosting,
the efficiency of the shocked pulsar wind in converting the spin-down luminosity
into non-thermal HE gamma-ray emission must be quite high, and would only weaken
the constraint for certain phases (see Section~\ref{sec:doppler}).

The acceleration and emission processes responsible for the high
efficiency in the conversion from spin-down luminosity to HE gamma-ray
emission are currently unknown.  IC emission from particles accelerated
beyond the light cylinder in the striped pulsar wind is one of the
proposed emission processes \citep{2011MNRAS.tmp.1193P}. However, when
applied to \psr, this process can only account for the low-flux state
before periastron, and is unable to explain the high luminosity detected
after the periastron passage without the invocation of addition seed
photon sources apart from the stellar companion. The extremely high
efficiency required for \ls, similar to the high-flux state of \psr,
indicates that this process might not be responsible for its HE
gamma-ray emission.

\subsection{Concluding remarks} 

The semi-analytic model presented here has allowed us to gain valuable
insight into the nature of \ls.  The assumptions described in
Section~\ref{sec:thermal} are chosen such that the resulting level of
thermal X-ray emission is fairly conservative and thus the upper limit
on the spin-down luminosity robust. 
We have shown that the study of the thermal emission from the shocked
stellar wind can be a powerful diagnostic tool for pulsar gamma-ray
binaries. Even for cases where no thermal features are observed in the
X-ray spectrum, it can be used to place constraints on important
properties of the system such as stellar mass-loss rate and pulsar
spin-down luminosity. The application of such a model to \ls\ allowed us
to conclude that, if hosting a non-accreting pulsar, this system has a
very efficient non-thermal mechanism, with 
$\lsd\sim(3\mbox{--}6)\times10^{36}\,\ergs$.

\acknowledgements 
We acknowledge support by the Spanish Ministerio de Ciencia e Innovaci\'on
(MICINN) under grants AYA2010-21782-C03-01 and FPA2010-22056-C06-02. V.Z.~was
supported by the Spanish Mi\-nis\-te\-rio de Edu\-ca\-ci\-\'on through FPU grant
AP2006-00077. V.Z.~and V.B.-R.~thank the Max Planck Institut f\"ur Kernphysik for its
kind hospitality and support.  
The research leading to these results has received funding from the European
Union Seventh Framework Programme (FP7/2007-2013) under grant agreement
PIEF-GA-2009-252463.
J.M.P.~acknowledges financial support from ICREA Academia.

\emph{Facilities:} \facility{XMM (EPIC)}

\bibliographystyle{apj}
\bibliography{ls-ads}

\end{document}